\documentclass[12pt,iopams,setstack,epsfig,epsf,float]{iopart}
\usepackage{cite}
\usepackage{epsf}
\usepackage{epsfig}
\usepackage{float}
\usepackage{amssymb}

\def\ep{{\epsilon}}

\def\om{{\omega}}

\def\g{{\bf{g}}}

\def\P{{\bf{P}}}

\def\d.c.{{\scriptscriptstyle d.c.}}
\def\beq{\begin{equation}}
\def\eeq{\end{equation}}
\def\beqa{\begin{eqnarray}}
\def\eeqa{\end{eqnarray}}

\def\g0{{\gamma_0}}

\eqnobysec
\begin{document}
\bibliographystyle{plain}
\input epsf

\title[Optical and transport properties of heavy fermions: theory and experiment]{
 Optical and transport properties of heavy fermions: theory compared to experiment.}

\author{N S Vidhyadhiraja and David E Logan}

\address{ University of Oxford, Physical and Theoretical Chemistry
Laboratory,\\ South Parks Rd, Oxford OX1~3QZ, UK}

\begin{abstract}

 Employing a local moment approach to the periodic Anderson model within the
framework of dynamical mean-field theory, direct comparison is made between
theory and experiment for the d.c.\ transport and optical conductivities of
paramagnetic heavy fermion and intermediate valence metals. Four materials,
exhibiting a diverse range of behaviour in their transport/optics, are analysed
in detail: $CeB_6$, $YbAl_3$, $CeAl_3$ and  $CeCoIn_5$. Good agreement between
theory and experiment is in general found, even quantitatively, and a mutually
consistent picture of transport and optics results.

\end{abstract}


\submitto{\JPCM}

\section{Introduction.}
\label{sec:intro}
Heavy electron materials have long been the subject of extensive 
investigation, for reviews see e.g.\ \cite{grew91,hews,aepp,fisk,taka,degi,rise}. 
Yet in many respects even their `normal' paramagnetic phase, 
be it metallic or insulating, has eluded a unified microscopic description 
on all experimentally relevant temperature ($T$) and/or frequency ($\om$) scales.
The canonical theoretical model here is of course the periodic Anderson model (PAM). 
Within the general framework of dynamical mean-field theory~\cite{voll,prus95,geor,gebh} we have developed in the preceding paper~\cite{vidh05} (here referred 
to as I) a non-perturbative local moment approach to paramagnetic metallic 
phases of the PAM, with a focus on d.c.\ transport and optics; following 
earlier work on  $T=0$ dynamics~\cite{vidh04} and on Kondo 
insulators~\cite{smit,vidh03}. The primary emphasis of I is the Kondo 
lattice regime relevant to strong correlated heavy fermion (HF) metals. 
Dynamics/transport on all relevant ($\om, T$)-scales are encompassed, 
from the low-energy behaviour characteristic of the lattice coherent
Fermi liquid, through incoherent effective single-impurity physics to non-universal
high-energy scales. The underlying theory is not however restricted to the Kondo
lattice regime, enabling it also to handle e.g.\ intermediate valence (IV) behaviour.

  The present paper is an attempt to provide at least a partial answer 
to the question: to what extent are the optical and d.c.\ transport 
properties of HF and related materials captured by the PAM and our theory 
for it? That clearly requires direct, quantitative comparison between 
theory and experiment, which is our purpose here. Specifically, 
we consider in detail the transport and optics of four materials:
$CeB_{6}$, $YbAl_{3}$, $CeAl_{3}$ and $CeCoIn_{5}$, all HF metals with the
exception of the IV compound $YbAl_{3}$, and exhibiting a diverse range of
behaviour in their transport and optical behaviour. The materials are 
analysed on a case by case basis in \S s 3-6, following a discussion (\S 2) 
of relevant issues involved in making the comparison; and we believe it 
fair to claim that the theory provides a striking account of experiment.

\section{Background issues}

  The Hamiltonian for the PAM is given by equation (2.1) of I, and
its physical content is simple: a single correlated $f$-level 
in each unit cell hybridizes locally to an uncorrelated conduction band.
The model is moreover specified by only four `bare'/material parameters
rendering it minimalist in terms of comparison to experiment --- the more 
so when one recalls that it encompasses regimes of behaviour as diverse as
strongly correlated heavy fermion metals and Kondo insulators, 
intermediate valence, and weak coupling. The dimensionless bare parameters are 
$U$, $V$, $\epsilon_{c}$ and $\epsilon_f$ (in units
of the conduction electron hopping, $t_{*} \equiv 1$), with
$U$ denoting the local  $f$-level interaction strength and $V$ the 
local one-electron hybridization coupling an $f$-level to the conduction band.
The energy of the local conduction orbital, $\epsilon_c$, determines
the centre of gravity of the free conduction band relative to the Fermi
level (and thereby the conduction band filling, $n_c$); and 
$\epsilon_f$ denotes the $f$-orbital energy. An equivalent parameter set
is $U$, $V$, $\epsilon_c$ and $\eta$, where $\eta = 1+2\epsilon_f/U$ specifies
the $f$-level asymmetry.

The non-interacting limit of the model ($U=0$) is certainly trivial.
But in that case  --- for \emph{all} $T$ ---  the d.c.\ resistivity of the metallic state vanishes, and the optical conductivity contains no absorption below the direct gap save for a $\delta$-function Drude contribution at $\om =0$, see I. That this behaviour bears scant comparison to experiment reflects the fact that
the essential physics is driven by scattering due to electron interactions. 
It is of course the latter, and the resultant many-body nature of the problem, 
that renders the PAM non-trivial.

In considering d.c.\ transport, the first requirement in comparing theory to 
experiment is thus to extract the contribution ($\rho_{\rm mag}^{exp}(T)$) 
to the measured resistivity ($\rho(T)$) that isolates the interaction 
contributions from those of phonons and static impurity scattering. 
This is given, ideally, by
\begin{displaymath}
\rho_{\rm mag}^{exp}(T) = \{\rho(T)-\rho(0)\} - \rho_{\rm ph}(T). 
\end{displaymath}
The first term removes the residual ($T=0$) resistivity, and hence
impurity scattering contribution on the assumption that the latter is
$T$-independent. The second removes the contribution from phonons;
usually taken in practice (as assumed in the following) to be the 
resistivity of the non-magnetic homologue compound with the magnetic ion 
$Ce$ (or $Yb$) replaced by $La$ (or $Lu$), on the assumption that interactions 
in the latter are negligible. For most systems the phonon contribution
is generally negligible for $T\lesssim 50K$ or so. 
 
  While this prescription is straightforward in 
principle, we first deal with a complication that can arise in practice. Experimentally it is the  resistance that is measured directly. To convert to a sample independent resistivity requires spatial dimensions to be known with reasonable precision. That does not pose a problem with large crystals, but may do for small samples. Examples arise in the literature where reported $\rho(T)$'s from different groups differ quite significantly; we encounter one such in the case of $CeCoIn_{5}$ considered in \S 6. Absolute resistivities would thus be related to
measured values by e.g.\ $a^{\prime}\rho(T)$ and $a^{\prime\prime}\rho_{\rm ph}(T)$, where $a^{\prime}$ and $a^{\prime\prime}$ denote experimental `mismatch' factors. 
For comparison
to theory (equation (2.2) below) we require however only the relative
factors $a^{\prime}/a^{\prime\prime}$, to which end we
replace the above by
\beq
\rho_{\rm mag}^{exp}(T) = a\{\rho(T)-\rho(0)\} - \rho_{\rm ph}(T)
\eeq
where $a=1$ in the ideal case (the majority of systems considered below).

  The experimental $\rho_{\rm mag}^{exp}(T)$ is to be compared to $\rho_{\rm mag}(T)$
arising from the theory of I for the PAM (where $\rho_{\rm mag}$ was denoted simply 
by $\rho$). The system is generically characterized by a low-energy coherence scale 
$\om_L = ZV^{2}$, with $Z$ the quasiparticle weight or inverse mass renormalization factor. This scale is a complicated function of the underlying bare parameters, see
e.g.\ \cite{vidh04} and refs therein. But in the strong coupling Kondo lattice regime 
$\om_L$ is exponentially small (because $Z$ is). In consequence, 
$\rho_{\rm mag}(T)$ exhibits scaling in terms of $\om_L$, i.e.\ is of form
\beq
\rho_{\rm mag}(T) = \alpha H\left(\frac{T}{\om_L}\right)
\eeq
with the temperature dependence encoded in $T/\om_L$, 
\emph{independent} of the interaction strength $U$ and hybridization $V$
($\alpha$ is a trivial overall scale factor, $\alpha \equiv 1/\sigma_{0}$ in
the notation of I); the scaling holding for (any) fixed $\epsilon_c$ 
and $\eta$. The scaling resistivity is moreover only weakly dependent on $\eta$, and for $T/\om_L \gtrsim 1-5$ or so is in fact independent of $\epsilon_c$, reflecting the crossover to incoherent effective single impurity physics (as detailed in I, see 
e.g.\ figure 8). This enables direct connection to experiment, via the extent 
to which the scaling form equation (2.2) captures the $T$-dependence of experimental
resistivities; and indeed also their pressure dependence, for although
$\om_L$ will change with pressure the scaling behaviour should remain intact 
(we consider this explicitly in the case of $CeAl_3$, \S 5).
Success in this regard also enables $\om_L$ to be determined 
directly, there being  no hope of calculating it `ab initio' with any meaningful accuracy. Finally, we add that comparison of theory/experiment proceeds along the same lines away from the asymptotic Kondo lattice regime, in dealing e.g.\ with intermediate valence materials or intermediate/weak coupling compounds. Here a full parameter set $\epsilon_c$, $\eta$, $U$ and $V$ must in general be specified; 
but $\rho_{\rm mag}(T)$ can always be cast in the form equation
(2.2), and appropriate comparison can be made.

  Although our comments above focus on static transport, a central purpose
of the paper is also to make direct comparison between experiment and theory 
for optics, on all experimentally relevant frequency and temperature scales.
Given prior analysis of d.c.\ transport, the underlying model/material parameters 
are known, either wholly or in part. There is then little room for manoeuvre; the resultant theory either captures the optical behaviour or not, providing a 
further and quite stringent test of the model's material applicability.

 The main physical effect omitted in the PAM itself is that of crystal electric 
fields (CEFs). The atomic levels of e.g.\ $Ce^{3+}$, $^{2}F_{\frac{5}{2}}$,
are generically split into three doublets, the excited levels lying
above the ground state by $\Delta_{1}$, $\Delta_{2}$. For sufficiently 
low $T$ only the lowest level matters, it being this alone the PAM seeks
to capture. Although the material specific $\Delta_{i}$ are usually 
larger than $\om_L$ (itself typically $\sim 10 -100 K$), they often
lie in the interval $\lesssim 300 K$. Their qualitative influence on d.c.\
transport is clear, for additional conduction channels are opened up
in accessing higher CEF levels with increasing $T$.
Quantitatively however, the effect is hard to gauge \emph{a priori}, its magnitude naturally depending on how effectively the higher CEF levels couple to the
conduction band. Where present and effective, we can expect to see it as
a decrease in the experimental resistivity below that predicted by the 1-channel
PAM, the onset of the deviation appearing at $T \sim \Delta_{1}$. We 
add here that while the role of CEF effects has been studied in the context 
of \emph{single}-impurity Anderson/Kondo models~\cite{kash,stro,corn72,yama,suzu}, with application to lattice-based systems for sufficiently high temperatures where lattice coherence can be neglected, we are not aware of corresponding work in the context of lattice-fermion models.

  Finally, we note that our comparison of theory/experiment 
for Kondo insulators in~\cite{vidh03} was
free from most of the considerations above. There, theory was compared directly
to the experimental $\rho(T)$, for several reasons. First, in 
contrast to the case of heavy fermion metals, resistivities of the non-magnetic 
homologues are sufficiently small compared to those of the Kondo insulating 
material that $\rho_{\rm ph}(T)$ in equation (2.1) can be neglected 
with impunity. That in turn means that any sample geometry factor $a \neq 1$ in
equation (2.1) can simply be absorbed into the overall scale factor $\alpha$  
(equation (2.2)), so any lack of precision in obtaining $\rho(T)$ is immaterial.
The role of impurities is also different in Kondo 
insulators. In metals this arises from static impurity scattering, presumed
to be $T$-independent and generating the finite residual resistivity
$\rho(0)$; which is thus subtracted out as in equation (2.1). In Kondo
insulators by contrast $\rho(T)$ diverges as $T \rightarrow 0$, reflecting
the insulating ground state. This occurs even in the presence of localized
impurity states in the $T=0$ insulating gap; which generate conduction by
variable-range hopping, generally operative over a narrow $T$-interval 
(e.g.\ $\lesssim 8 K$ for $SmB_{6}$~\cite{gors,vidh03}), and whose net effect on
$\rho(T)$ at temperatures above this interval (where direct comparison is
made to experiment) is usually sufficiently small to be neglected.

 We turn now to comparison with experiment for the metallic heavy fermion
and intermediate valence materials considered.

\section{$CeB_6$}
Among the rare-earth hexaborides, the Kondo insulator $SmB_6$
and the heavy fermion compound $CeB_6$ have been investigated for 
many years~\cite{grew91,rise,sato,kimu,naka,good,marc}.
The former was considered in our recent work~\cite{vidh03} on the particle-hole
symmetric PAM. There it was shown that a single low-energy (indirect gap) scale underlies the temperature dependence of both the static and optical conductivity,
and (with minimal input of bare material parameters) the frequency dependence 
of the optics as well. Here we consider its metallic counterpart $CeB_6$, likewise 
a cubic system~\cite{sato}.
At the lowest temperatures, various antiferromagnetic phase transitions occur between
$T\sim 1.6-3.3K$~\cite{sato,naka,marc}; above (and below) which the paramagnetic phase arises, `Phase-I' for $T > 3.3K$.

The relative ease with which large, clean single crystals of 
$CeB_6$ can be grown (see e.g.\ \cite{marc}) has been a 
motivating factor in its investigation.
The d.c.\ resistivity has been measured by several groups (e.g.\ \cite{sato,marc}) and since the sample quality is in general good, the residual resistivity 
is very small. In addition, the relatively large crystal sizes
imply small errors in sample geometry, hence the factor $a$ in
equations (2.1) can safely be assumed to be 1.
The phonon contribution to the resistivity is as usual taken as the
resistivity of the non-magnetic homologue $LaB_6$~\cite{sato}, the latter
having the same lattice structure with similar lattice parameters
and phonon dispersion as $CeB_{6}$ (from inelastic neutron scattering~\cite{sato}).
The experimental magnetic resistivity $\rho_{\rm mag}^{exp}(T)$ is then readily obtained~\cite{sato}, and shown in figure 1 (we add that the
phonon contribution kicks in only for $T\gtrsim 50K$, so 
the magnetic resistivity essentially coincides with the
raw resistivity at lower temperatures). A small kink arises in 
$\rho_{\rm mag}^{exp}(T)$ at $T\sim 3.3K$, reflecting the transition from  paramagnetic phase I to phase II~\cite{sato,marc} mentioned above. The experimental resistivity is seen to exhiCoverletter~  fig4.eps  iopart10.clo  paper.tex       ReplyLetter~
bit the
classic `shape' for a strongly correlated HF metal,
increasing from zero at $T=0$ and going through a coherence peak at
$T\sim 4K$, before decreasing through a small log-linear regime (similar 
to that shown in figure 6(a) of I) to a shallow minimum at $T\sim 375K$;
increasing thereafter at higher temperatures where it shows conventional 
metallic behaviour.

  To compare $\rho_{\rm mag}^{exp}(T)$ to theory, first recall from I that the asymptotic scaling resistivity $\rho_{\rm mag}(T)$ is a universal function of
$T/\om_L$ for
fixed $\epsilon_c$ and $\eta$ (being independent of the local interaction $U$ and hybridization $V$); see e.g.\ figure 8 of I. For a chosen $(\epsilon_c,\eta)$ the scaling $\rho_{\rm mag}(T)$ is then straightforwardly superposed onto 
$\rho_{\rm mag}^{exp}(T)$ with an appropriate rescaling of the temperature and
resistivity axes, thus enabling the low-energy scale $\om_L$ to be determined. The asymptotic scaling $\rho_{\rm mag}(T)$ is shown in figure 1 (dotted line) for a moderate $\epsilon_c =0.5$ and $\eta =0$, with the resultant coherence scale 
thereby found to be $\om_L \simeq 5.5K$.
It is seen to capture the experimental resistivity 
up to $T \simeq 100K$, but above this it deviates below experiment, continuing 
as it must to decrease monotonically (see figure 8 of I) and hence lacking the minimum occurring experimentally for $CeB_{6}$ at  $T_{\rm min} \simeq 375K$, i.e.\ 
$\tilde{T}_{\rm min} = T_{\rm min}/\om_L \simeq 70$. 

  As discussed in I (figure 8 inset), the latter behaviour is 
physically natural and readily encompassed 
theoretically. No real HF material is in the universal scaling
regime `for ever' --- it must be exited sooner or later with increasing $T$. Deviation of $\rho_{\rm mag}(T)$ from its asymptotic scaling form  at sufficiently high temperatures signifies the onset of non-universality; and the location of the non-universal minimum in $\rho_{\rm mag}(T)$ provides an opportunity to identify the
ratio $U/V^{2}$ of effective bare material parameters, which will be helpful in making a prediction for the $\omega$-dependence of the optical conductivity 
$\sigma(\om;T)$. Specifically, for the chosen $\epsilon_c =0.5, \eta =0$, we find 
that the theoretical $\rho_{\rm mag}(T)$ indeed has a minimum at 
$\tilde{T}_{\rm min} \simeq 70$ (as in experiment) for $U/V^{2} \simeq 12$.
This is illustrated in figure 1, where we show the resultant theoretical
$\rho_{\rm mag}(T)$ for $U=4.75, V^{2} =0.4$ (solid curve) and
$U=2.45, V^{2}=0.2$ (dashed curve). The two $\rho_{\rm mag}(T)$'s barely
differ from the asymptotic scaling resistivity for $T\lesssim 100K$, are essentially coincident with each other across the entire $T$-range, and each possesses
a weak minimum at $T \simeq 375K$. 
Except naturally for a small neighbourhood around the low-temperature phase 
transitions (which the theory does not seek to address), the resultant theoretical 
$\rho_{\rm mag}(T)$ is seen to be in rather good agreement with experiment, 
as evident further from the inset to figure 1 where the corresponding d.c.\
conductivity $\sigma_{\rm mag}(T)$ is shown. A CEF excitation is known to
occur at $\sim 2.5 meV$ ($30 K$)~\cite{zirn} (with a second lying at a much
higher energy, $46 meV$~\cite{zirn}); but as judged from the above comparison
this appears to play a minor role in the d.c.\ transport itself.

\begin{figure}[h]
\epsfxsize=300pt
\centering
{\mbox{\epsffile{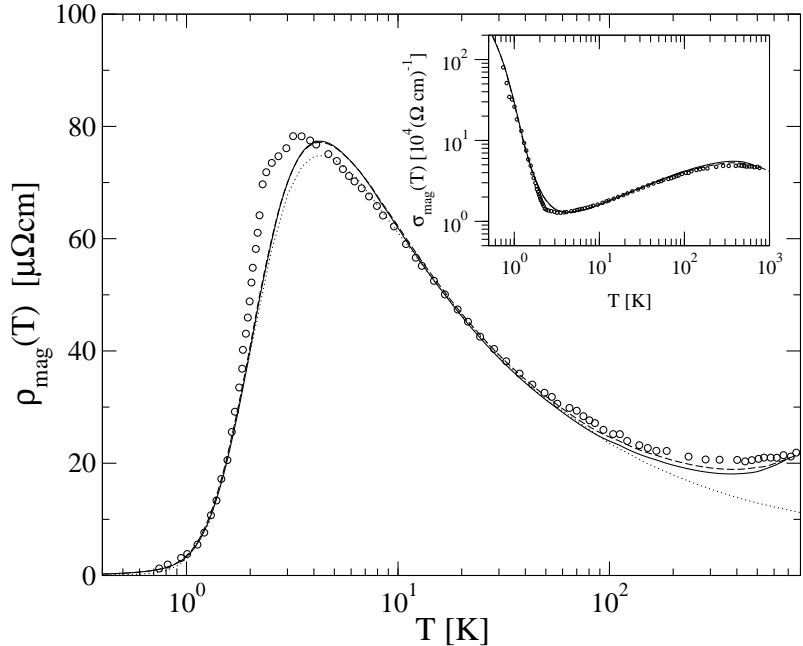}}}
\caption{Comparison of experimental $\rho_{\rm mag}^{exp}(T)$ (circles) for $CeB_6$ 
 ~\cite{sato} to theory, on a log-linear scale. The solid curve shows 
$\rho_{\rm mag}(T)$ for $U=4.75, V^2=0.4$, while the dashed curve
is for $U=2.45, V^2=0.2$; both theory sets have common $\ep_c=0.5, \eta=0$
and the same $\om_L=5.5K$. The dotted curve shows the asymptotic scaling
resistivity.
Inset: the d.c.\ conductivity $\sigma_{\rm mag}(T)\equiv 1/\rho_{\rm mag}(T)$ on a 
log-log scale.
}
\end{figure}

  To determine the optical conductivity on all $\om$-scales, including non-universal
frequencies, requires $U$ and $V^{2}$ to be separately specified as discussed in
I. That is clearly
not provided by the above analysis, but a reasoned prediction for $\sigma(\om;T)$ 
may nonetheless be made. For fixed $U/V^{2} \simeq 12$ as above, we find that varying 
$U$ across the range $\sim 2.5-5$ produces only a modest change in both the $\om$- and $T$-dependence of $\sigma(\om;T)$, leading in particular to a direct gap absorption lying in the interval $\sim 200-300 ~cm^{-1}$; as well as a quasiparticle weight $Z$ on the order of $10^{-2}$ that is consistent with the effective mass $(m^{*}/m_{e}\equiv 
)$ $m^* \simeq 100$ deduced from the
specific heat coefficient $\gamma=250 ~mJ\,mol^{-1} K^{-2}$ measured for Phase-I
of $CeB_{6}$~\cite{sato,marc}. In otherwords, on the assumption that the reasonably wide $U$-range above encompasses the behaviour of $CeB_{6}$, the optics are relatively insensitive to the precise value of $U$. 
Coverletter~  fig4.eps  iopart10.clo  paper.tex       ReplyLetter~

  To that end we show in figure 2 the predicted optical conductivity for $U=2.45$, $V^2=0.2$, with the $\om/\om_L$-dependence of the theoretical $\sigma(\om;T)$ converted to $\om$ in $cm^{-1}$ using $\om_L =5.5K
(\equiv 3.8cm^{-1})$ deduced from the above analysis of the d.c.\ resistivity;
and with $\sigma(\om;T)$ shown for a range of temperatures from
$1.1K$ to $660K$. As $\om \rightarrow 0$, the $T$-dependence of the dynamical
conductivity follows the d.c.\ values shown in the inset to figure 1. At the lowest
temperature $T=1.1 K \simeq \om_L/5$ an emergent low-frequency Drude absorption on frequency scales $\om \lesssim 1~cm^{-1}$ is evident in $\sigma(\om;T)$ (see also figure 9 of I),
separated by a clear optical pseudogap centred on $\om \approx 10~cm^{-1}$ from 
a strong direct gap absorption centred on $\om \sim 250~cm^{-1}$. The Drude absorption is rapidly suppressed on increasing $T$, being all but dead by
$T \sim \om_L =5.5K$; while the optical pseudogap is progressively `filled in' on temperature scales
set by $\om_L$. The direct gap absorption is largely unaffected by temperature until
$T\gtrsim (5-10)\om_L$ or so; but is significantly eroded by $T \sim 110K$ and in essence destroyed by room temperature.

\begin{figure}[h]
\epsfxsize=320pt
\centering
{\mbox{\epsffile{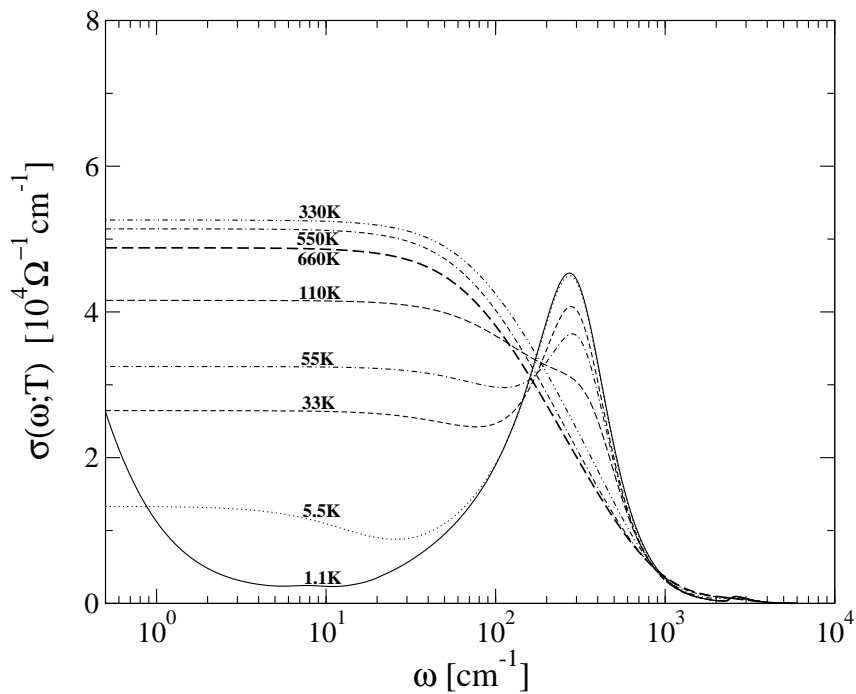}}}
\caption{ Predicted optical conductivity of $CeB_6$: $\sigma(\om;T)$ vs.\ $\om$ in 
$cm^{-1}$ at temperatures $T=1.1K, 5.5K, 33K, 55K, 110K, 330K, 550K$ and $660K$.
}
\end{figure}

The optical conductivity of $CeB_{6}$ has in fact been measured by
Kimura \etal ~\cite{kimu}, but only in the frequency interval 
$50meV \simeq 400cm^{-1}$ to $10eV\simeq 8\times 10^4 cm^{-1}$, and at 
a temperature of $300K$.
No strong direct gap absorption was observed, itself suggesting that the absorption
occurs at frequencies on the order of $\om \sim 200~cm^{-1} \simeq 300K$ or less;
instead a broad, featureless and monotonically decreasing spectral lineshape was found. This is of course consistent with the prediction from figure 2 above that the direct gap peak is almost completely washed away at room temperature. In order to observe non-trivial $\om$- and $T$-dependence in the optical conductivity 
of $CeB_{6}$, we thus suggest the frequency domain be extended down to 
$\sim 10~cm^{-1}$ (or lower to observe the Drude absorption), and that experiments 
be performed at considerably lower temperatures such as those shown in figure 2.

\section{$YbAl_3$}

  Our main focus in I and \cite{vidh04} has been the strongly correlated heavy
fermion regime where the $f$-level $\epsilon_{f} \ll 0$ lies well below the
Fermi level, with $\epsilon_f +U \gg 0$ well above it, such that the 
$f$-electrons are essentially localized, $n_{f} \rightarrow 1$.
The underlying local moment approach is not however restricted to the Kondo
lattice regime, and in particular can also readily handle intermediate 
valence (IV) behaviour. In this case, depletion for example of $n_f$ from unity reflects the fact that $\epsilon_f$ lies relatively close to the Fermi level, such
that the appropriate $\eta = 1+2\epsilon_f/U$ regime is $\eta \approx 1$.

  The compound $YbAl_3$, which crystallizes in a simple cubic $Cu_3Au$-type structure and  does not order magnetically~\cite{ohar}, provides a prime example of IV behaviour. Figure 3 shows the experimental d.c.\ resistivity of 
$YbAl_3$, and its non-magnetic homologue $LuAl_3$, at ambient pressure (data 
from \cite{ohar} with the tiny residual resistivity subtracted). The resultant
magnetic resistivity $\rho_{\rm mag}^{exp}(T)$ is also shown in figure 3, 
obtained simply by
subtracting the resistivity of $LuAl_3$ from that of $YbAl_3$ (the samples~\cite{ohar}
are high quality single crystals, hence the factor $a$ in equation (2.1) is
taken as unity). $\rho_{\rm mag}^{exp}(T)$ is seen to increase
monotonically with $T$, lacking the coherence peak seen in strongly correlated HF materials. This behaviour is characteristic of IV~\cite{schw91}, as too is e.g.\ 
the low/moderate effective mass in the range $m^*\sim 15-30$ inferred from 
dHvA~\cite{ebih}, optical~\cite{okam} and specific heat~\cite{corn02} measurements; 
and the $Yb$ mean valence $z_{v} \equiv 2+n_{f}$ is estimated experimentally as 
$z_{v}\sim 2.65-2.8$\cite{corn02,suga}.

In the strongly correlated HF regime, physical properties such as $\rho_{\rm mag}(T)$
exhibit scaling as a function of $T/\omega_L$, independent of the interaction $U$ and hybridization $V$ as detailed in I; occurring formally for all $T/\omega_L$ in the asymptotic strong coupling limit, and in material practice over a significant albeit 
naturally finite $T/\om_L$ range as 
seen above for $CeB_{6}$ (and in~\cite{vidh03} for Kondo insulators).  
This is \emph{not} by contrast the case in the IV regime, and neither is it to be expected. Here the `full set' of bare/material parameters $\eta, \epsilon_c, U$ and $V$
must be specified. To compare theory to experiment for $YbAl_{3}$ we
consider a moderate $\epsilon_c = 0.5$ with $\eta =1.2, U=4.9$ and $V^2 =0.8$;
corresponding to a modest interaction strength $U/\pi \Delta_{0} = 1.4$ (with
$\Delta_{0} =\pi V^2 \rho_{0}(-\epsilon_c)$ as in I) and an $\epsilon_f =0.49$ close
to the Fermi level ($\epsilon_f/\Delta_{0} = 0.44$). The resultant $f$-band filling
is found to be $n_{f} \simeq 0.65$ ($z_{v} \simeq 2.65$), consistent with the mixed valence nature of $YbAl_{3}$; and a quasiparticle weight $Z \simeq 0.05$ is found,
implying an effective mass $m^{*} \simeq 20$ that is likewise consistent with
experiment as above.

\begin{figure}[t]
\epsfxsize=300pt
\centering
{\mbox{\epsffile{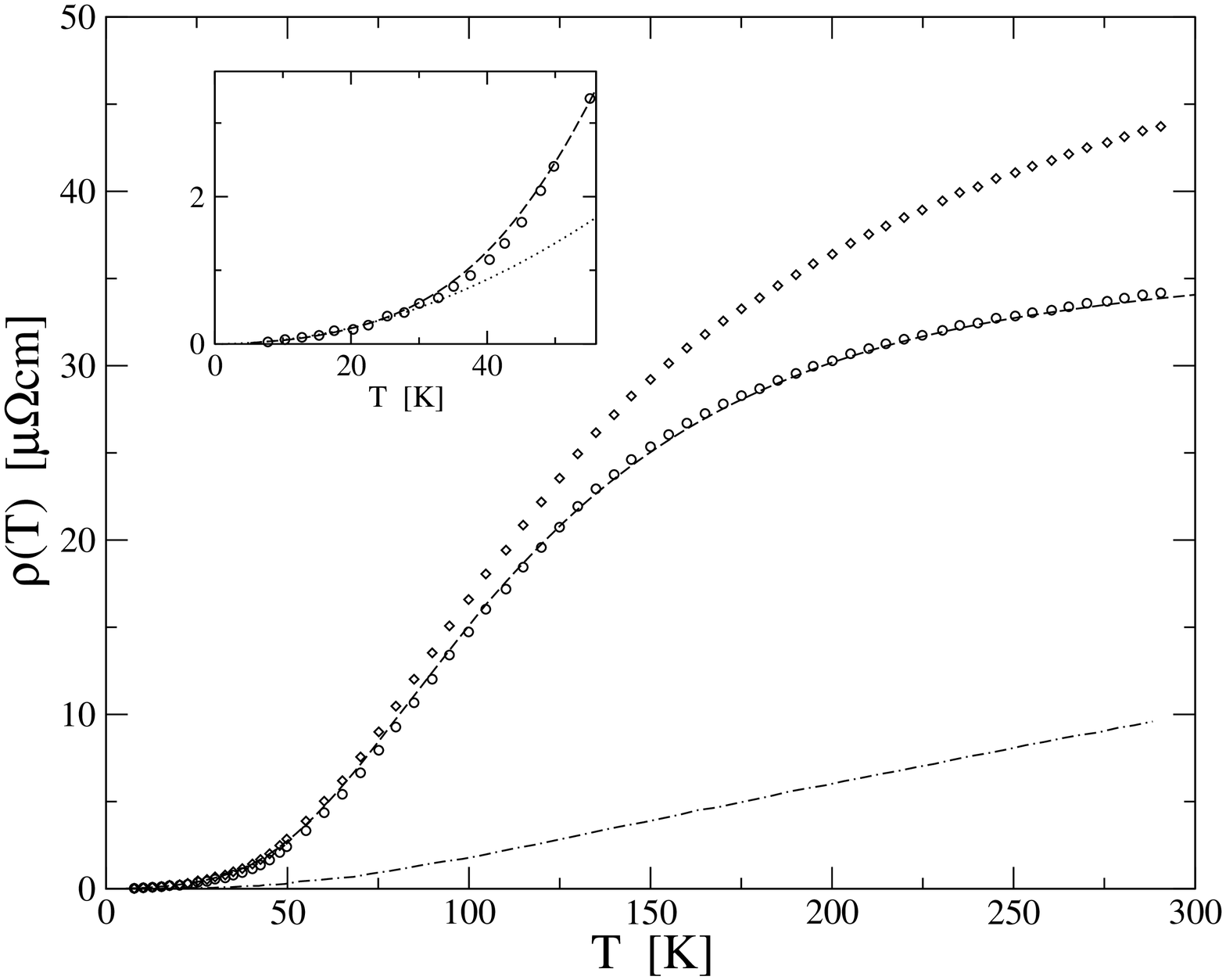}}}
\caption{
The d.c.\ resistivity of $YbAl_3$ (diamonds) and $LuAl_3$ (point-dash line), 
from \cite{ohar} with the small residual resistivity subtracted.
The experimental $\rho_{\rm mag}^{exp}(T)$ (circles) is obtained by subtracting 
$\rho(T)-\rho(0)$ for $LuAl_3$ from that for $YbAl_3$.  The theoretical
$\rho_{\rm mag}(T)$ (dashed line) is obtained for $\ep_c=0.5, \eta =1.2, U = 4.9, 
V^2=0.8$, and superimposed on $\rho_{\rm mag}^{exp}(T)$ with $\om_L=254K$.
The level of agreement between theory and experiment is clear. Inset: the low
temperature behaviour; including (dotted line) a fit to the asymptotic Fermi liquid 
form $\rho_{\rm mag}(T) \propto T^2$, which is seen to persist in both experiment
and theory up to $T \sim 30 K$.
}
\end{figure}

The theoretical d.c.\ resistivity $\rho_{\rm mag}(T)$ vs.\ $T/\om_L$ 
for these parameters  
has been determined, and superposed onto the experimental $\rho_{\rm mag}^{exp}(T)$
in the usual way. The resultant low-energy scale is found thereby to be
$\om_L =254K$, and comparison between theory and experiment is shown in figure 3.
The agreement is clearly excellent, for all temperatures. The inset to figure 3
shows the low temperature behaviour on an expanded scale, together with a fit (dotted
line) to the $T \rightarrow 0$ Fermi liquid form $\rho_{\rm mag}(T) \propto T^2$. As 
known from experiment~\cite{corn02}, and seen also in the present theory, this asymptotic form is seen to persist up to a temperature `$T_{FL}$'$\sim 30 K$ that is an order of magnitude lower than $\om_L \sim 254 K$; and about which fact we make two brief comments. First, to emphasise that this arises naturally here within a  theoretical approach to the periodic Anderson model itself (with no appeal e.g.\ to calculations based upon a single-impurity Anderson model~\cite{corn02}). Second,
the quantitative distinction between `$T_{FL}$' and $\om_L$ is in our view to be anticipated; for $\om_L$ is the natural low-energy scale in the problem, and since 
$\rho(T) \propto T^2$ is the \emph{asymptotic} $T\rightarrow 0$ behaviour, we 
would as such expect it to arise only for $T \ll \om_L$.

\begin{figure}[t]
\epsfxsize=300pt
\centering
{\mbox{\epsffile{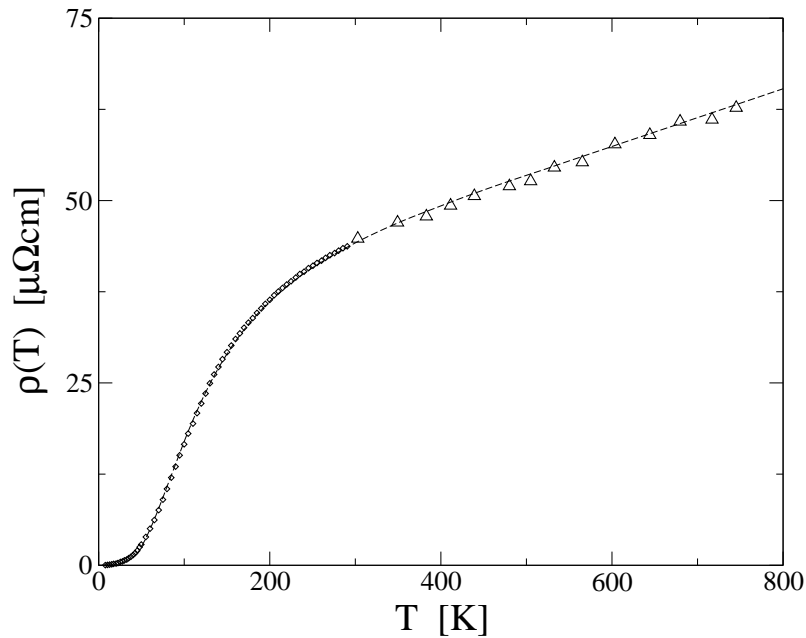}}}
\caption{
Theoretical $\rho(T)$ for $YbAl_{3}$ (dashed line) up to $800 K$, compared 
to the experimental results of~\cite{ohar} (diamonds) up to $T=300 K$ and 
of~\cite{rowe} from $300-750 K$ (triangles), as detailed in text.
}
\end{figure}

  While the experimental data shown in figure 3 extend up to $T \simeq 300 K$, the
theoretical $\rho_{\rm mag}(T)$ may be used to predict the full $\rho(T)$ for
$YbAl_{3}$ over a much larger temperature interval. To that end we simply calculate 
$\rho_{\rm mag}(T)$ at higher $T$, and add to it the resistivity of 
$LuAl_{3}$ representing the phonon contribution, itself extended to higher $T$ by linear extrapolation of the $Lu$ data in figure 3 (point-dash line, and which
extrapolation is clearly warranted). The resultant $\rho(T)$ is shown
in figure 4 out to $T =800K$; and we note that it continues to increase monotonically above $300 K$, precluding as such the occurrence of a coherence maximum at a $T$ in excess of that shown in figure 3. Experimental
results for $\rho(T)$ up to $T \simeq 750 K$ have in fact been reported~\cite{rowe}. 
This data does not appear to be quite as clean as that of~\cite{ohar}
in the interval $50 K \lesssim T < 300 K$, but for $T \lesssim 50 K$ the $\rho(T)$ from~\cite{rowe} collapses very well onto that of~\cite{ohar} (considered above) with 
an overall $\rho$-axis rescaling factor $a =1.2$ (equation (2.1)). Taking this
$a$, the resultant $\rho(T)$ from~\cite{rowe} is shown in figure 4 in the
temperature interval $300 -750 K$; and is seen to agree well with the theoretical
result.

  We turn now to the optical conductivity of $YbAl_{3}$, which has only
recently been measured at infrared frequencies and below~\cite{okam}.
This is reproduced in the top panel of figure 5, from which three key features 
are evident~\cite{okam}: the low-frequency Drude response characteristic of the
free carriers; a depleted pseudogap occurring at low temperatures (at 
$\om \sim 20~meV$) and flanked on its right by a shoulder at $\om \sim 50-60~meV$; followed by the broad, strong direct gap 
(or mid-infrared, mIR) peak centred near $\om \sim 250 meV$.

\begin{figure}[h]Coverletter~  fig4.eps  iopart10.clo  paper.tex       ReplyLetter~

\epsfxsize=360pt
\centering
{\mbox{\epsffile{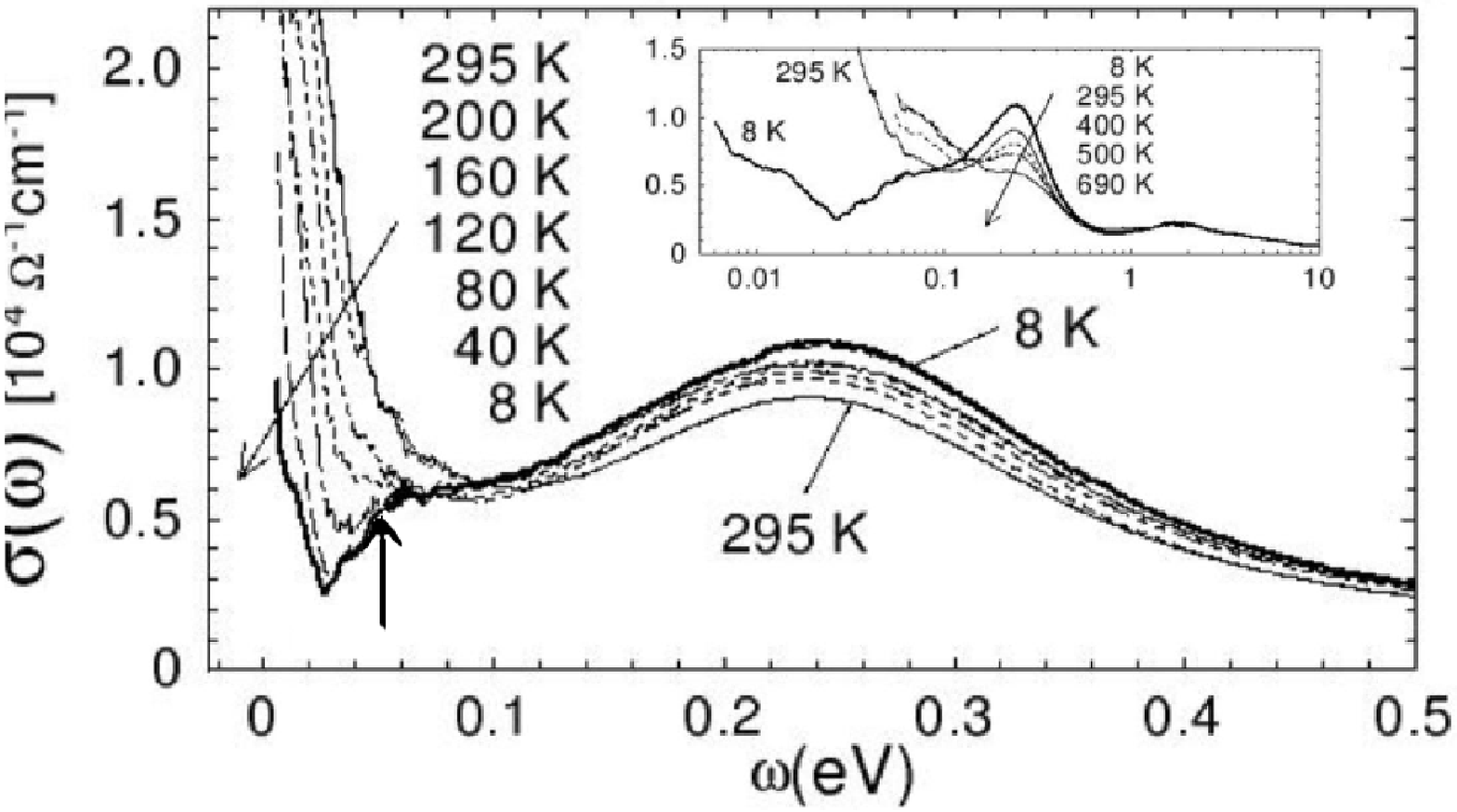}}}
\epsfxsize=340pt
\centering
{\mbox{\epsffile{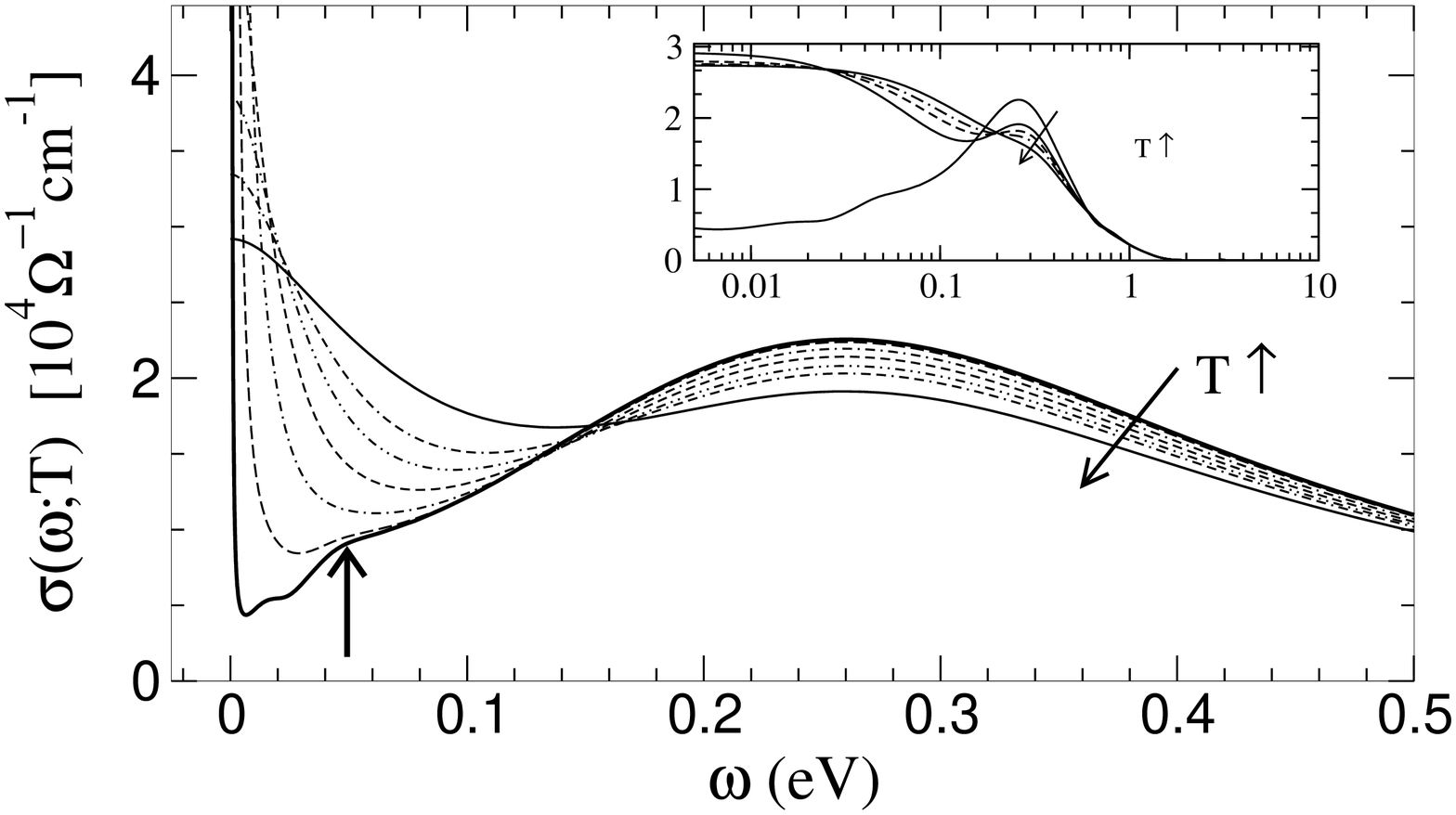}}}
\caption{
Top panel: experimental optical conductivity of $YbAl_3$~\cite{okam}.
Bottom panel: theoretical $\sigma(\om;T)$ for the same parameters as in 
figure 3; at the same temperatures and on the same frequency scale as in 
experiment. The vertical arrow in both panels indicates the position of
the theoretically predicted shoulder, at $\om \simeq 2\om_L \sim 50~meV$.
The insets show the thermal evolution of $\sigma(\om;T)$ for
higher temperatures.
The overall experimental lineshape including the pseudogap, shoulder and 
mIR peak, as well as their thermal evolution, is well reproduced by theory. 
}
\end{figure}

The bottom panel in figure 5 shows the theoretical optical conductivity
at the same temperatures and on the same frequency scale as experiment, obtained
using the same bare parameters employed in figures 3,4, and with $\om_L =254 K$ as deduced above; in otherwords with no additional input other than that inferred
from the d.c.\ transport comparison.
Barring a mismatch in the relative intensity of the direct gap peak on which
we comment below, the agreement between theory and experiment is seen to be rather good in terms of the pseudogap structure, the shoulder, the position and width of 
the mIR peak, and the thermal evolution of the optical conductivity. The shoulder
in particular merits comment, being a distinct spectral feature that is thermally
destroyed with increasing temperature~\cite{okam}.  It has been speculated 
in~\cite{okam} (not unreasonably) that its origin may lie outside the scope 
of the PAM and/or be material specific. The present results however suggest
to the contrary. Theoretically, we find that the existence of the shoulder shown in figure 5 is not specific to the particular set of bare material parameters considered, but rather characteristic of IV in general terms --- a clear optical shoulder arising in the vicinity of $\om \simeq 2-3 \om_L$ as the bare parameters are varied over quite a significant range. Its origins reflect the underlying behaviour of the $f$-electron self-energy, and we do not therefore have a simple physical explanation for it; but neither do we doubt its generic occurrence. For the case of $YbAl_{3}$, with $\om_L \sim 250 K$ deduced from d.c.\ transport as above, the shoulder thus lies at $\om \sim 50~meV$ as indicated by the vertical arrows in figure 5 and agreeing rather well with experiment.

Finally, we comment on the mismatch in the vertical (intensity) axes in figure 5. The
theoretical $\sigma(\om;T)$ represents of course the conductivity in the
absence of phonons, which inevitably introduces a certain mismatch between
theory and experiment as $\om \rightarrow 0$ at temperatures high enough for
the phonons to kick in; but the phonon background would not extend up to mIR frequencies, so the intensity of the direct gap peak in theory and experiment
should agree. The fact that it does not
could obviously mean that a different bare parameter set may be more appropriate 
than that identified here. Alternatively, the issue may be experimental.
The $\sigma(\om;T)$ spectra are obtained~\cite{okam}
via Kramers-Kr\"onig transformation of the total reflectance  
between $7meV$ and $35eV$, with a Hagen-Rubens formula used for low-energy extrapolation; and with seCoverletter~  fig4.eps  iopart10.clo  paper.tex       ReplyLetter~
veral spectrometer sources employed in different frequency intervals, which require matching using appropriate constant factors (see e.g.\ \cite{awas}). This is intricate, and there is undoubtedly the possibility of error 
in determining the absolute intensity of the direct gap.
A surface impedance probe such as the one used in \cite{awas} might be able to resolve the matter; while if the issue is not experimental then a more extensive scan of the bare parameter space is required. That notwithstanding, however, the present theory does appear to provide a remarkably consistent description of both transport and optics in $YbAl_{3}$.

\section{$CeAl_3$}

The classic system $CeAl_{3}$ has long been subject to extensive investigation,
see e.g.\ \cite{grew91,rise03,awas,kaga,oomi,andr,gore99,jacc88,jacc85}.
In contrast to its $Yb$ cousin considered above, $CeAl_{3}$ is a
prototypical heavy fermion material; as attested for example by the
large specific heat coefficient 
$\gamma \sim 1.4\times 10^{3}~ mJmol^{-1}K^{-2}$~\cite{andr}
and corresponding effective mass $m^{*} \sim 700$~\cite{andr,awas}.

  A helpful starting point for comparison of theory to experiment is a rough knowledge
of the parameter regime to which the system belongs. In the case of $CeAl_3$
we can glean such information from the unusual behaviour of the
experimental optical conductivity~\cite{awas}, which is shown in the top panel
of figure 7 below.
Typical features found in optical lineshapes of HF systems are the Drude peak
at low frequencies, followed by a pseudogap, with a strong direct gap (or mIR)
absorption at higher frequencies (see e.g.\ figures 2,5,10). In the case of $CeAl_3$
however, while the Drude peak is clearly evident at low $T$, a distinct 
pseudogap and the mIR peak are absent. In fact there is little frequency
dependence beyond the Drude absorption range.

  From our theoretical work in I we know that increasing $\epsilon_c$, 
and hence reducing the conduction band filling $n_c$, acts to diminish 
the pseudogap (see inset to figure 10 of I); relatedly, it also tends to 
suppress the direct gap absorption, suggesting that a suitably
large $\epsilon_c$ is required for $CeAl_3$. However this by itself is not 
sufficient: a large $\epsilon_c$, but with a modest hybridization $V$ 
between the $f$-levels and the conduction band, can still 
give rise to a distinct mIR absorption. A suitably large hybridization
thus also seems necessary to suppress strongly the direct gap absorption. 
Unusually large CEF parameters found in inelastic neutron scattering
experiments have in fact also been attributed to a large hybridization 
$V$~\cite{gore99}, supporting this inference. And from the large 
effective mass mentioned above, we know the system is in the 
strong correlated Kondo lattice regime, requiring a significant interaction
strength $U$.

  The picture of $CeAl_3$ thus suggested is of a system with
low conduction band filling $n_c$ (large $\epsilon_c$), a significant 
hybridization $V$, and strong local interactions. We have investigated this 
regime in some detail, and for comparison to transport and optical experiments
on $CeAl_3$ will consider explicitly the following material parameters: 
$\epsilon_c = 1.5$ and $\eta =0$ (results are quite insensitive to $\eta$ in the 
Kondo lattice regime), together with $V^{2}=1.4$ and $U=6.9$ (or
$U/\pi\Delta_{0} \simeq 8.4 \gg 1$ signifying strong correlations,
where $\Delta_{0} = \pi V^{2}\rho_{0}(-\epsilon_c)$
as in I). Additional specification of $U$ and $V^{2}$ is of course required in
order to consider the optical conductivity on all frequency scales,
whereas $\epsilon_c$ and $\eta$ alone suffice to determine the d.c.\ 
resistivity in the scaling regime.
With these parameters we find a quasiparticle weight $Z \simeq 1.6 \times 10^{-3}$
and hence an effective mass $m^{*} \simeq 625$, in good agreement with 
$m^{*} \simeq 690$ deduced from specific heat measurements ~\cite{andr,awas}.
The resultant conduction band filling is $n_c = 0.17$; while the $f$-level
occupancy $n_{f} = 1.0$, consistent with the firmly trivalent ($z_{v} \equiv 4-n_f$)
nature of the $Ce$ ion expected in the HF regime.
\begin{figure}[h]
\epsfxsize=300ptCoverletter~  fig4.eps  iopart10.clo  paper.tex       ReplyLetter~

\centering
{\mbox{\epsffile{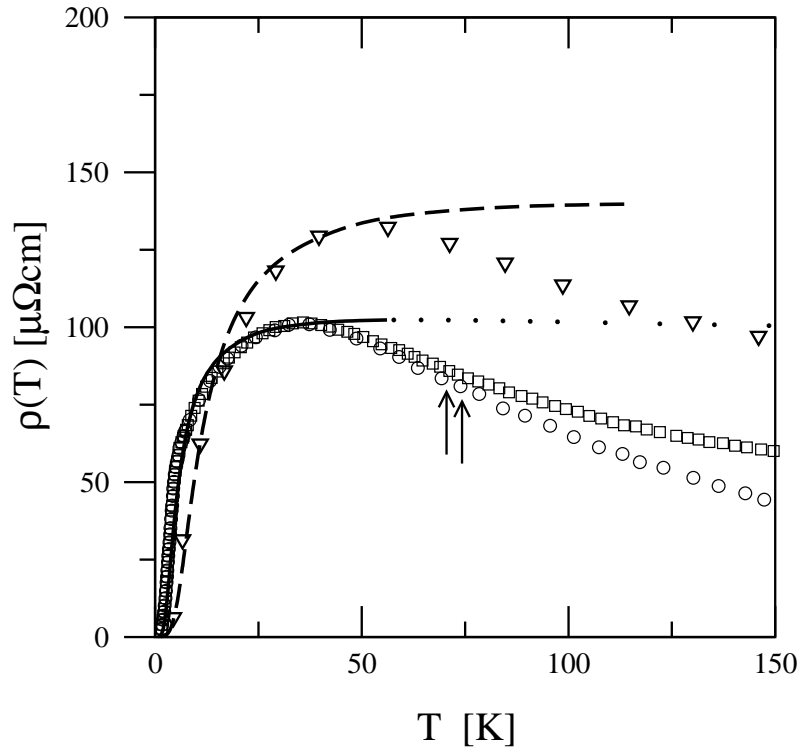}}}
\caption{
d.c.\ resistivity of $CeAl_3$ at ambient pressure (squares, from~\cite{kaga} 
with the residual resistivity subtracted); and the corresponding magnetic resistivity 
$\rho_{\rm mag}^{exp}(T)$ (circles)~\cite{kaga}. The theoretical
$\rho_{\rm mag}(T)$ for $\epsilon_c=1.5$, $\eta =0$ is shown by the solid/dotted
line, the resultant coherence scale being $\om_L =33K$. The agreement between theory and experiment is good for $T\lesssim 45K$, beyond which deviations 
naturally occur due to the presence of two crystal field split levels at 
$6.1 meV ~(71K)$ and $6.4 meV ~(75K)$~\cite{gore99}, marked by arrows. 
Triangles denote $\rho_{\rm mag}^{exp}(T)$ for 
$CeAl_3$ at a higher pressure $P= 0.4~GPa$~\cite{oomi}. The dashed line
shows the same theoretical curve as used for comparison to the ambient pressure data, but with simple rescaling of the axes, leading to $\om_L(0.4~GPa)=71K$. 
}
\end{figure}
 
 We first consider d.c.\ transport measurements. These have been obtained by
several groups~\cite{andr,kaga,oomi,awas,jacc88}, which we find in general concur well on subtraction of appropriate residual resistivities (so 
$a=1$ is taken in equation (2.1)). We choose to compare explicitly to the 
$\rho(T)$ data of~\cite{kaga}, which is shown in figure 6 (squares) at ambient 
pressure with the residual resistivity subtracted out; together with the corresponding 
$\rho_{\rm mag}^{exp}(T)$ (circles) obtained by further subtracting the resistivity 
of $LaAl_3$. Figure 6 also shows corresponding results for $\rho_{\rm mag}^{exp}(T)$
obtained at a pressure $P =0.4~GPa$~\cite{oomi} (triangles), using the same
sample as in~\cite{kaga}.
  The theoretical $\rho_{\rm mag}(T)$ is calculated and scaled onto the ambient
pressure $\rho_{\rm mag}^{exp}(T)$ in the usual way, leading thereby to a coherence
scale of $\om_L \simeq 33K$. From figure 6, comparison between theory and
experiment is seen to be good up to $T \simeq 45K$, beyond which the experimental
$\rho_{\rm mag}^{exp}(T)$ drops much more rapidly with further increasing $T$.
This is natural, for inelastic neutron scattering experiments~\cite{gore99}
show two higher crystal field levels occuring at almost the same energy,
$6.1meV ~(70.8K)$ and $6.4meV ~(74.2K)$ above the ground state. Marked in
figure 6, these are
accessed thermally as $T$ approaches $\sim 70K$, and provided they couple
effectively to the conduction band (as appears to be the case here) the two 
additional conduction channels diminish the resistivity significantly.

  We now comment briefly on the high pressure magnetic resistivity shown in
figure 6. The coherence scale $\om_L = ZV^{2}$ itself naturally varies with
pressure, but in the strongly correlated Kondo lattice regime the $T$-dependence
of the magnetic resistivity should depend universally on $T/\om_L$ alone
(\S 2). The same theoretical $\rho_{\rm mag}(T)$ employed for comparison
to the ambient pressure data should thus, with mere rescaling of the axes, 
account for $\rho_{\rm mag}^{exp}(T)$ at $P=0.4~GPa$ (again up to the
temperature at which the CEF excitations kick in). That indeed it does is shown in
figure 6 (dashed line); the resultant coherence scale being found to be 
$\om_L(P=0.4~GPa) \simeq 71K$.

\begin{figure}[t]
\epsfxsize=350pt 
\centering
{\mbox{\epsffile{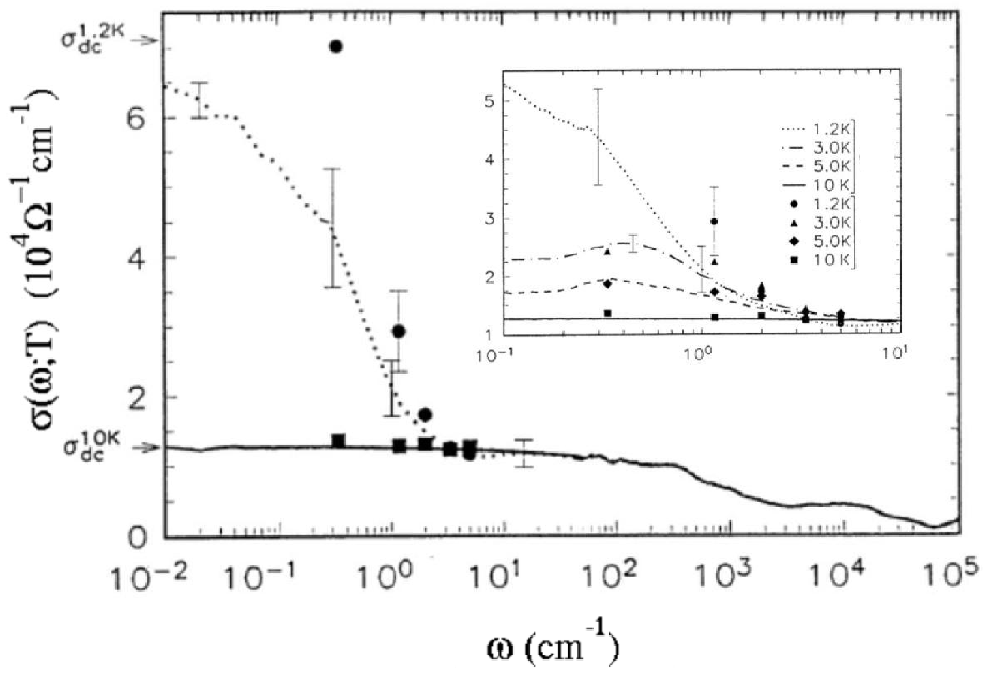}}}
\epsfxsize=343pt
\centering
{\mbox{\epsffile{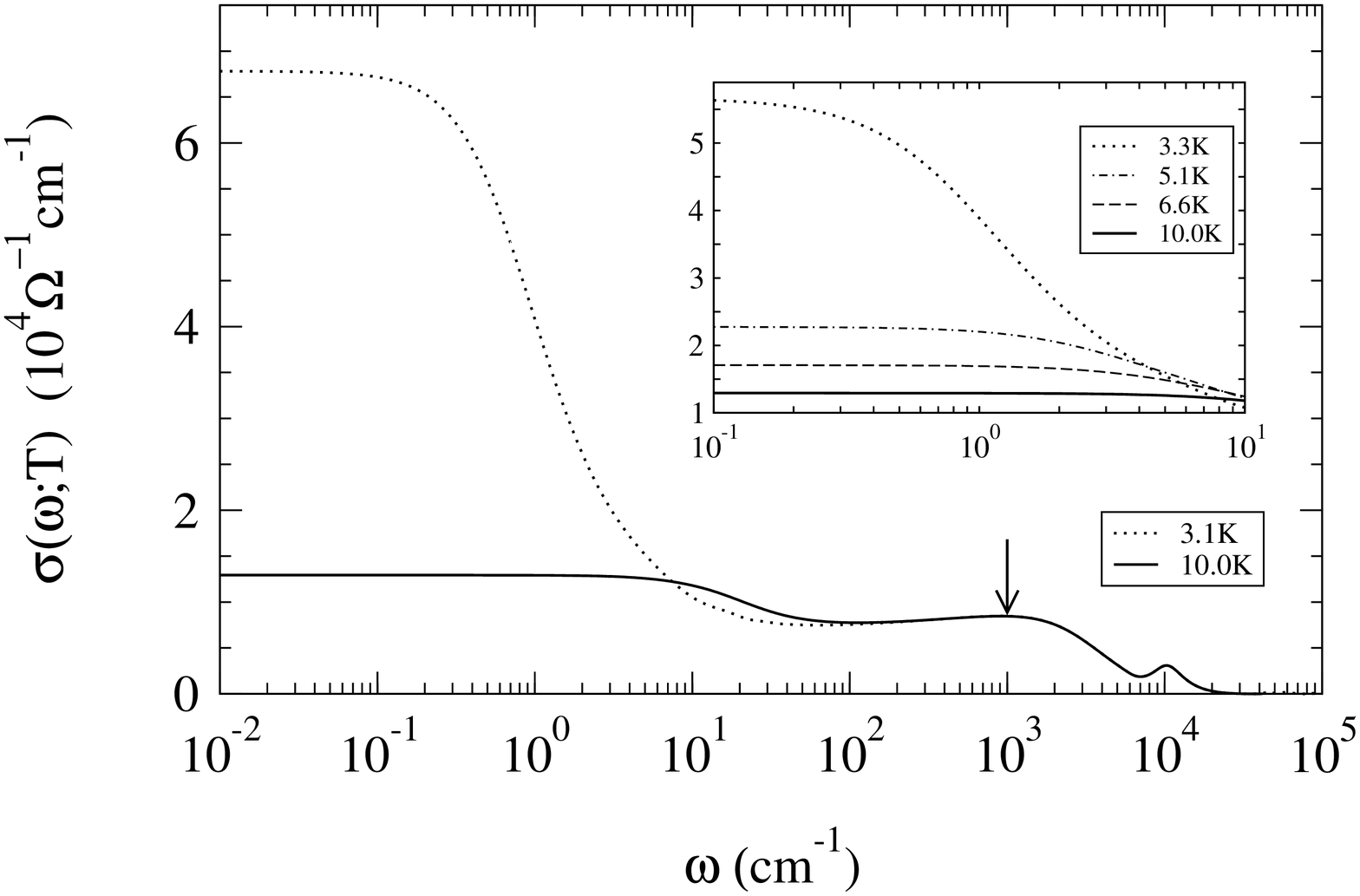}}}
\caption{Top panel: experimental optical conductivity of $CeAl_3$ at
the temperatures indicated, from~\cite{awas}. 
Bottom panel: theoretical $\sigma(\om;T)$. Discussion in text.
The vertical arrow denotes the location of the direct gap, where a mIR peak 
is usually found for other HF systems.
}
\end{figure}

  Finally we turn to the optical conductivity, the experimental
results~\cite{awas} shown in figure 7 (top panel) being obtained from
reflectance spectroscopy (lines) and surface impedance measurements (points).
The bottom panel shows the theoretical $\sigma(\om;T)$ obtained with the
bare parameters specified above, converted to $cm^{-1}$ using $\om_L = 33K$
inferred above from static transport. The theory evidently captures the
unusual optical characteristics of $CeAl_3$, and agreement with experiment
is seen to be rather good.
A strong Drude peak, and a very shallow pseudogap, are seen at the lowest 
temperatures in theory and experiment; the Drude absorption collapsing on a temperature scale of $\sim 10K$, above which 
very little $T$ dependence is found across the entire frequency range.
Most significantly, no distinct direct gap/mIR peak arises, the spectrum
being largely featureless (the small feature visible at 
$\sim 10^{4} cm^{-1}$ in the theoretical $\sigma(\om;T)$ occurs on the 
effective bandwidth scale, and its intensity diminishes further with 
increasing $U$). The nominal location of the direct gap itself can however be 
determined theoretically frCoverletter~  fig4.eps  iopart10.clo  paper.tex       ReplyLetter~
om the renormalized band structure underlying
the present theory (as considered in I, see figure 11). It is found to
lie at $\om \sim 10^{3} cm^{-1}$ as marked in figure 7, albeit that no
sharp absorption occurs in its vicinity.

\section{$CeCoIn_5$}
  This recently discovered~\cite{petr01}, moderately heavy fermion compound
crystallizes in a tetragonal structure consisting of alternate layers of $CeIn_3$ and $CoIn_2$. It superconducts below $T_c \simeq 2.3K$ 
(the highest transition temperature of all known HF systems at ambient
pressure~\cite{petr01}), and is paramagnetic for $T>T_c$.
The experimental resistivities $\rho(T)$ determined by four different 
groups~\cite{naka02,nick,petr01,sing} are shown in figure 8. That they differ 
widely presumably reflects intrinsic difficulties in measuring the dimensions of
relatively small samples. Their basic equivalence is however seen by taking 
the results of one as a reference and rescaling the $y$-axis for
each of the remaining data sets. With this, as shown in the inset to figure 8, 
all four resistivities collapse to essentially common form (that of~\cite{nick} deviating just slightly at higher $T$).
These differences are nonetheless potentially significant when comparing to theory, for which the magnetic contribution $\rho_{\rm mag}^{exp}(T)$ is required,
obtained as in equation (2.1) by subtracting 
the resistivity of the non-magnetic $LaCoIn_{5}$.
The results in figure 8
show that the $a$ factor in equation (2.1)
--- the relative weight of $\rho(T)$ compared to that for the 
non-magnetic homologue ---  
could vary by a factor of up to four or so, and is not therefore known with confidence.

\begin{figure}[h]
\epsfxsize=300pt
\centering
{\mbox{\epsffile{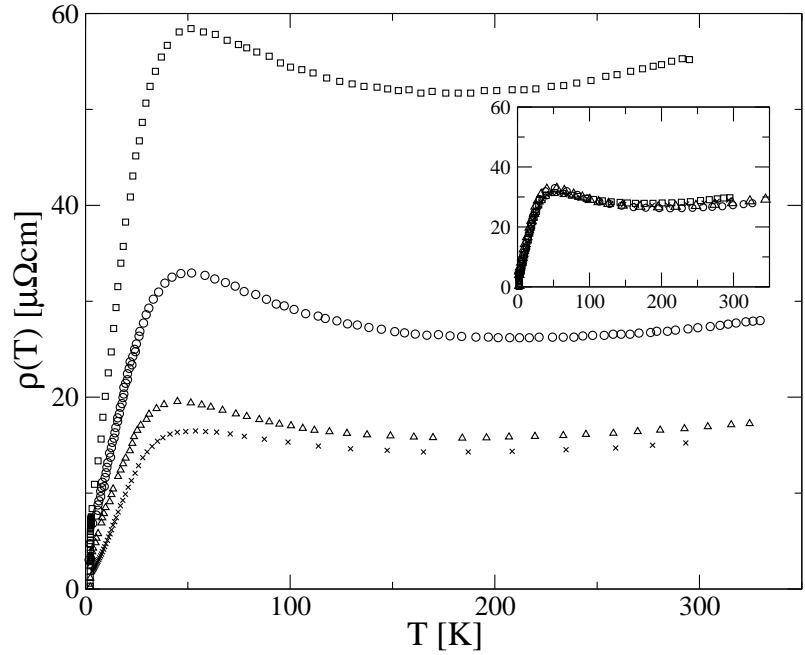}}}
\caption{Experimental resistivity of $CeCoIn_{5}$ measured by different groups: 
Ref~\cite{naka02} (circles), Ref~\cite{nick} (squares), Ref~\cite{petr01} (triangles)
and Ref~\cite{sing} (crosses). Inset: showing collapse of experimental
resistivities to common form on rescaling the $y$-axis alone.
}
\end{figure}

  Figure 9 shows the experimental $\rho_{\rm mag}^{exp}(T)$ (open circles)
determined in~\cite{naka02} by subtracting the resistivity of $LaCoIn_{5}$ from that for $CeCoIn_{5}$; corresponding as such to $a=1$ (or equivalently, if e.g.\ 
$\rho(T)$ from~\cite{sing} had been used instead, to an $a$ of $\sim 2$). 
\begin{figure}[h]
\epsfxsize=300pt
\centering
{\mbox{\epsffile{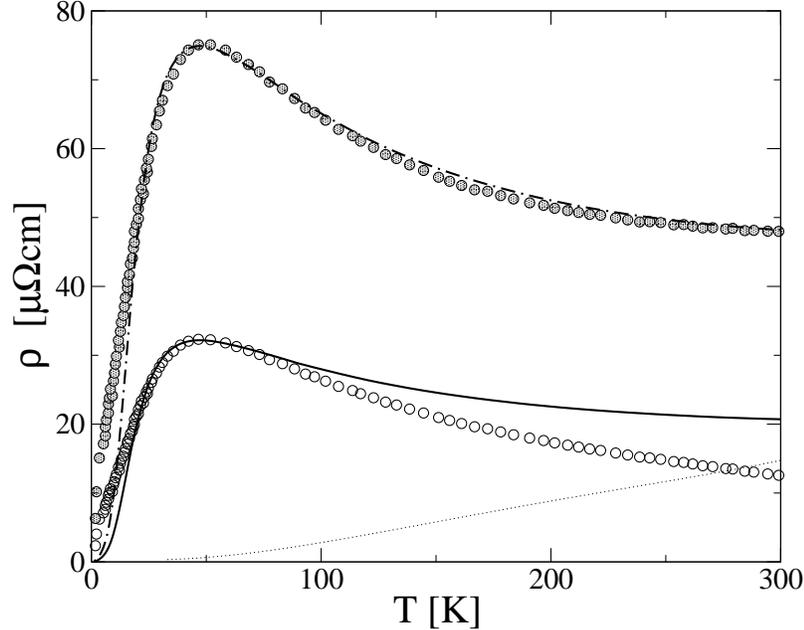}}}
\caption{
$CeCoIn_{5}$. Open circles denote the experimental $\rho_{\rm mag}^{exp}(T)$
from~\cite{naka02}, corresponding to $a=1$ in equation (2.1) (dotted line shows
the resistivity of $LaCoIn_{5}$). The solid line shows the theoretical
$\rho_{\rm mag}(T)$ for the bare parameters described in text, with coherence
scale $\om_L = 60K$. Filled circles denote an experimental $\rho_{\rm mag}^{exp}(T)$
obtained with $a=2.3$ instead; the corresponding theoretical $\rho_{\rm mag}(T)$
(again with $\om_L =60K$) is now shown as a point-dash line. Full discussion in text.
}
\end{figure}
To compare to theory we consider the parameters $\epsilon_c =0.5$ and $\eta =0$, with $U=3.75$ and $V^{2}=0.8$ (corresponding to an intermediate coupling 
strength $U/\pi\Delta_{0} \simeq 1$). The resultant theoretical 
$\rho_{\rm mag}(T)$ is compared to experiment in figure 9 (solid line), 
yielding a coherence scale of $\om_L = 60K$. It matches 
$\rho_{\rm mag}^{exp}(T)$ for $15 K \lesssim T \lesssim 100 K$,
the deviation at low temperatures presumably reflecting the approach
to the superconducting state. The deviation above $T \simeq 100 K$
also appears natural, since a direct determination of the CEF energy 
level scheme from inelastic neutron scattering~\cite{baue} shows an
excited level at $8.6 ~meV$ or $\sim 100 K$ (with a second at a much 
higher energy, $24.4 ~meV$), the extra conduction channel acting to
diminish $\rho_{\rm mag}^{exp}(T)$ more rapidly than the $1$-channel theory.

  Our guess is that the latter inference is correct, at least qualitatively.
A degree of caution is however required, since this 
is sensitive to a change in the value of $a$. To illustrate that, 
figure 9 also shows a new $\rho_{\rm mag}^{exp}(T)$ (filled circles) obtained with 
$a =2.3$ in equation (2.1). The theoretical $\rho_{\rm mag}(T)$ with the same coherence scale $\om_L = 60 K$, but with the overall $y$-axis naturally increased by factor of $2.3$, now describes $\rho_{\rm mag}^{exp}(T)$ very well for essentially all $T \gtrsim 15 K$. The quantitative influence of the extra conduction channel 
can thus be assessed with confidence only if the relative values of the resistivity
for the magnetic and non-magnetic compounds are known accurately;
although we add that the inferred low temperature coherence scale $\om_L=60K$
is not itself sensitive to $a$. 
\begin{figure}[t]
\epsfxsize=350pt
\centering
{\mbox{\epsffile{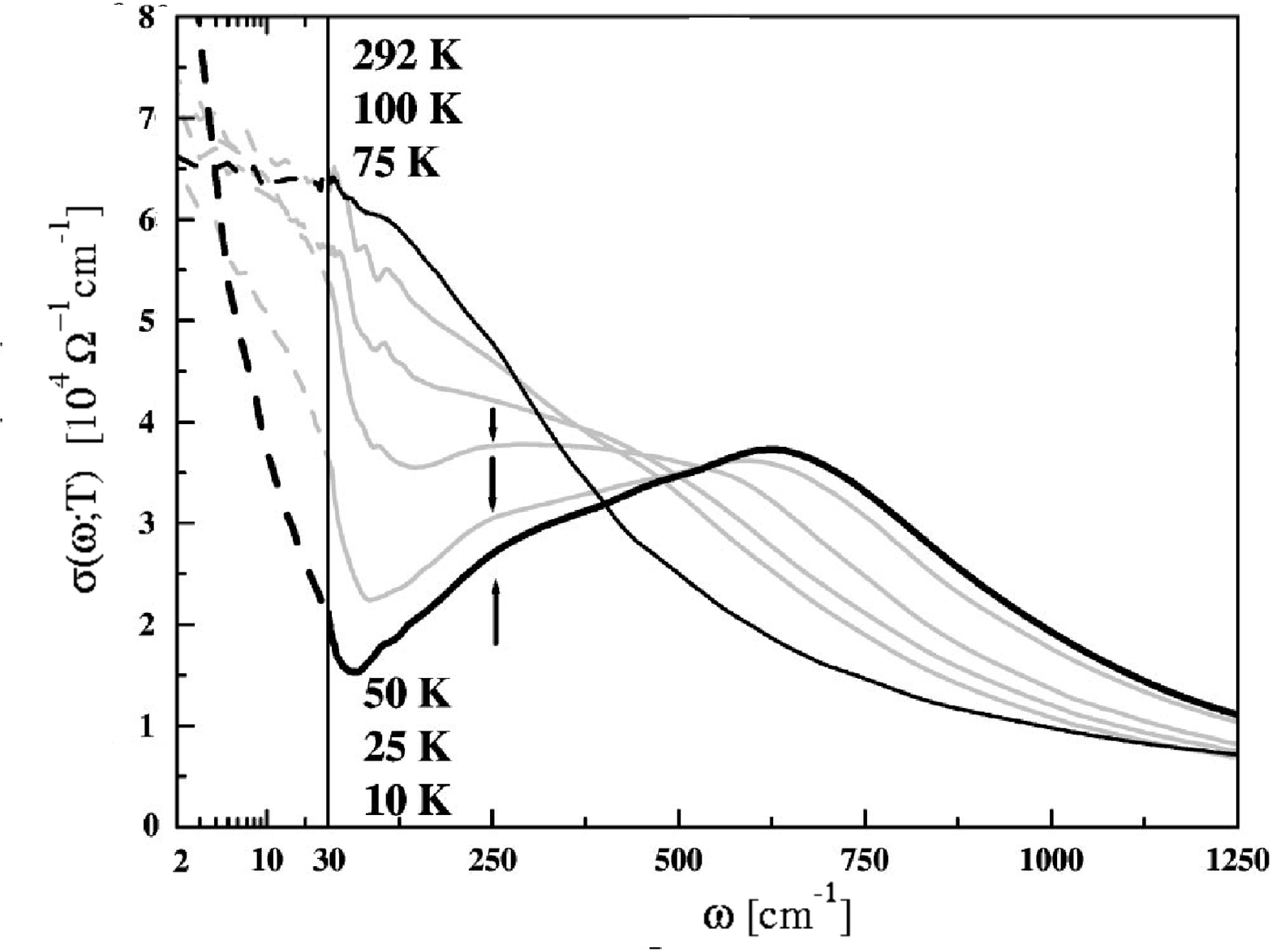}}}
\epsfxsize=330pt
\centering
{\mbox{\epsffile{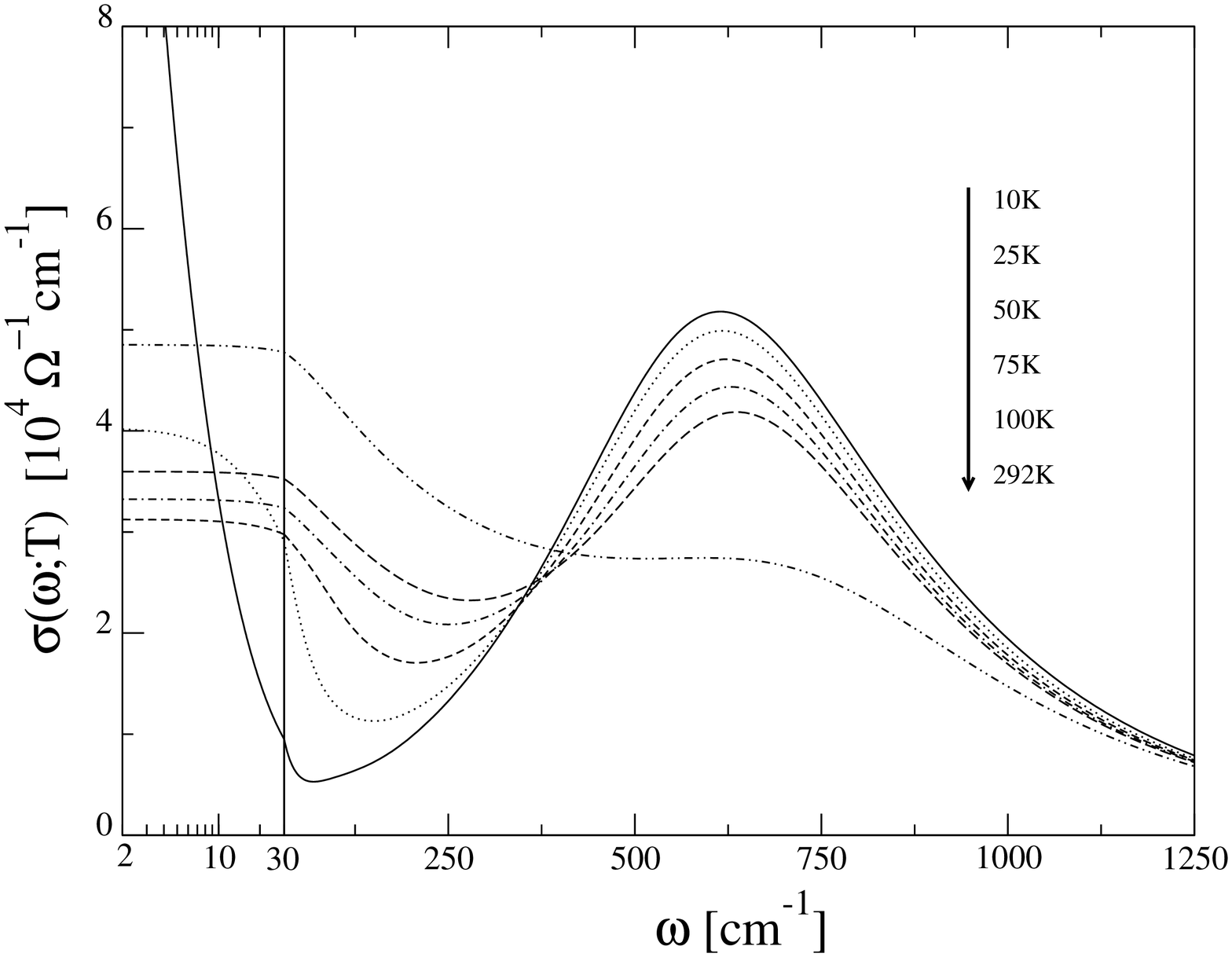}}}
\caption{Top panel: experimental optical conductivity of $CeCoIn_5$
at various temperatures~\cite{sing}. Bottom panel: corresponding theoretical  
$\sigma(\om;T)$ obtained for the parameter set $\epsilon_c =0.5, \eta =0$,
$U=3.75$ and $V^2=0.8$, with $\om_L=60K$ ($\sim 42 cm^{-1}$). In both cases the
frequency axis is logarithmic up to $\om =30cm^{-1}$ and linear thereafter
(separated by a vertical line). Full discussion in text.
}
\end{figure}

  We turn now to the optical conductivity. Experimental results 
from reflectivity measurements~\cite{sing} are shown in figure 10 (top panel),
with the $\om$-dependence on a log scale up to $\om = 30cm^{-1}$ and a linear scale thereafter (separated by a vertical line). For $\om < 30 cm^{-1}$ the 
results are extrapolated (dashed lines) towards the d.c.\ limit~\cite{sing}, as required for the Kramers-Kr\"onig analysis of the reflectivity that leads to the 
experimental $\sigma(\om;T)$ shown. The theoretical $\sigma(\om;T)$ is shown
in the lower panel, for the bare parameters specified above, with $\om_L =60K$.

The comparison between theory and experiment is at least qualitatively
satisfactory, albeit not as good as for $YbAl_{3}$ or $CeAl_{3}$ (perhaps
unsurprisingly given the quasi-2D nature of $CeCoIn_{5}$).
The pseudogap at $T= 10K$ is seen to lie at $\sim 65 cm^{-1}$,
and as in experiment shifts gradually to higher $\om$ with increasing temperature and
fills up progressively --- on the scale of $\sim 1-2 \om_L$ or $\sim 60-120 K$.
The direct gap absorption in theory and experiment lies at 
$\Delta_{\rm dir} \sim 600 cm^{-1}$;
and starts to lose spectral weight significantly for $T\gtrsim 50K$ or so,
i.e.\ on the scale of $\omega_L$ itself.
This suggests that $CeCoIn_{5}$ is not in the strong coupling,
Kondo lattice regime; since in strong coupling the direct gap/mIR absorption is 
significantly eroded for temperatures approaching the order of the direct gap itself 
($\sim 600 cm^{-1}$ or $\sim 900K$ in the present case), see e.g.\ figure 12 of I.
For $CeCoIn_{5}$ by contrast, significant thermal erosion is seen to occur 
for $T \gtrsim 50 K \sim \Delta_{\rm dir}/20$, which behaviour is typical of
intermediate coupling strengths. That also appears consistent with dHvA
measurements~\cite{hall}, which yield a moderate effective mass $m^{*}$ in the
range $\simeq 10-20$.

  Two further points should be mentioned. First, the 
$\om \lesssim 30 cm^{-1}$ values of the theoretical $\sigma(\om;T)$ are 
clearly lower than the experimental extrapolations. As $\om \rightarrow 0$, the 
latter extrapolate to the d.c.\ conductivity obtained in~\cite{sing}. 
The d.c.\ limit of the theoretical $\sigma(0;T)$ by contrast gives the 
$1/\rho_{\rm mag}(T)$ shown in figure 9 (solid line), and aside from 
the $T=292 K$ case this agrees well with the experimental 
d.c.\ conductivity of~\cite{naka02} --- which as seen from figure 8
differs by a factor of two or so from that of~\cite{sing}.
The issue here appears largely to be experimental, reflecting the 
significant difference between the resistivities of~\cite{naka02} 
and~\cite{sing}. Second, the vertical arrows 
at $\om \simeq 250 cm^{-1}$ in the experimental $\sigma(\om;T)$ of 
figure 10 indicate weak additional absorption that has been 
ascribed~\cite{sing} to a Holstein band due to coupling to a bosonic
mode. This is not of course included in
the present theory, which thus shows somewhat less absorption in the 
region.

\section{Conclusion}

We have here employed a local moment approach to the periodic Anderson 
lattice developed in I, to make direct comparison
to d.c.\ transport and optical conductivities of $CeB_{6}$, $YbAl_{3}$,
$CeAl_{3}$ and $CeCoIn_{5}$. The $Yb$ compound is a representative 
intermediate valence material, and the others typify heavy
fermion behaviour, from the strongly correlated Kondo lattice regime
appropriate to $CeAl_{3}$ and $CeB_{6}$ to what we believe is the
somewhat weaker coupling case of $CeCoIn_{5}$. 
In broad terms more or less all characteristic features of the optics 
and transport of these materials are captured; the natural exception, 
omitted from the model itself, being crystal field effects which may 
(or may not) show up in the experimental resistivity as a reduction 
below 1-channel behaviour at suitably high temperatures. The theory 
in general performs rather well quantitatively, and also captures
notable features specific to individual systems --- for example 
the existence of a low-frequency shoulder observed in the optics
of $YbAl_{3}$~\cite{okam}, or the absence of any significant 
direct gap/mIR absorption in $CeAl_{3}$~\cite{awas}.

  Minimalist though it is the underlying model, and theory for it, 
thus appear to provide quite a comprehensive and successful description of 
experiment. This we attribute in no small part both to the dominance 
of the local electron scattering inherent to the model
itself, and the need to provide an adequate theoretical description of
such on all experimentally relevant frequency and temperature scales.

\ack 
We are grateful to the following for permission to qoute their
experimental results: L.Degiorgi, H.Okamura, G.Oomi and F. Steglich.
We express our thanks to the EPSRC for supporting this research.

\end{document}